\documentclass[11pt]{article}
\usepackage{amssymb,amsfonts,amsmath,graphicx,hyperref}
\input{tcilatex}
\begin{document}

\title{Ricci Flow and Entropy Model for Avascular Tumor Growth and Decay Control}
\author{Tijana T. Ivancevic}\date{}
\maketitle

\begin{abstract}
Prediction and control of cancer invasion is a vital problem in
medical science. This paper proposes a modern geometric
Ricci--flow and entropy based model for control of avascular multicellular tumor spheroid growth and decay. As a tumor growth/decay control tool, a monoclonal antibody therapy is proposed.\\

\end{abstract}

\tableofcontents

\section{Introduction}

Cancer is one of the main causes of morbidity and mortality in the world.
There are several different stages in the growth of a tumor before it
becomes so large that it causes the patient to die or reduces permanently
their quality of life. Developed countries are investing large sums of money
into cancer research in order to find cures and improve existing treatments.
In comparison to molecular biology, cell biology, and drug delivery
research, mathematics has so far contributed relatively little to the area
\cite{Nature}.

On the other hand, consider the vital problem of \emph{prediction}
and \emph{control/prevention} of some natural disaster (e.g., a
hurricane). The role of science in dealing with a phenomenon/treat
like this can be depicted as the following OUPC feedback--loop
(see \cite{GaneshSprBig}):
\begin{equation*}
\begin{array}{ccccccc}
Observation & \longrightarrow  & Understanding & \longrightarrow &
Prediction &
\longrightarrow  & Control \\
\uparrow  &  &  &  &  &  & \downarrow \\
&\longleftarrow & \longleftarrow & \longleftarrow & \longleftarrow
& \longleftarrow
\end{array}%
\end{equation*}
with the following four components/phases:
\begin{enumerate}
\item $Observation$, i.e., monitoring a phenomenon in case, using
experimental sensing/measuring methods (e.g., orbital satellite
imaging). This phase produces measurement data that could be
fitted as graphs of analytical functions.

\item $Understanding$, in the form of geometric pattern
recognition, i.e., recognizing the turbulent patterns of
spatio--temporal chaotic behavior of the approaching hurricane, in
terms of geometric objects (e.g., tensor-- and spinor--fields).
This phase recognizes the observation graphs as cross--sections of
some jet bundles, thus representing the validity criterion for the
observation phase.

\item $Prediction$: when, where and how will the hurricane
strike?\newline Now, common, inductive approach here means fitting
a statistical model into empirical satellite data. However, we
know that this works only for a very short time in the future, as
extrapolation is not a valid predictive procedure, even if
(adaptive) extended Kalman filter is used. Instead, we suggest a
deductive approach of fitting some data into a well--defined
dynamical model. This means formulating a dynamical system on
configuration and phase--space manifolds, which incorporates all
previously recognized turbulent patterns of the hurricane's
spatio--temporal behavior. Once a valid dynamical model is
formulated, the necessary empirical satellite data would include
system parameters, initial and boundary conditions. So, this would
be a pattern--driven modelling of the hurricane, rather than blind
data--driven statistical modelling. This phase is the validity
criterion for the understanding phase.

\item $Control$: this is the final stage of manipulating the
hurricane to prevent the destruction. If we have already
formulated a valid geometric--pattern--based dynamical model, this
task can be relatively easily accomplished, as
\begin{equation*}
Control\;System\;=\;Dynamical\;System\;+\;Controller.
\end{equation*} So, here the problem is to design a feedback
controller/compensator for the dynamical model. This phase is the
validity criterion for the prediction phase.
\end{enumerate}

\begin{figure}[h]
\centerline{\includegraphics[width=5cm]{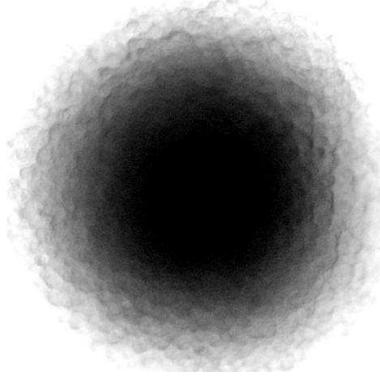}} \caption{An example
of multicellular tumor spheroid (MTS): a differential interference
contrast image.} \label{MTS}
\end{figure}

Since there are three distinct stages to cancer development:
avascular, vascular, and metastatic -- researchers often
concentrate their efforts on answering specific OUPC--related
questions on each of these stages (see \cite{Roose}). In
particular, as some tumor cell lines grown in vitro form
3--dimensional (3D, for short) spherical aggregates, the relative
cheapness and ease of in vitro
experiments in comparison to animal experiments has made 3D \emph{%
multicellular tumor spheroids} (MTS, see Figure \ref{MTS}, as well
as e.g., Figure 6 in \cite{Roose}) very popular in vitro model
system of avascular tumors\footnote{%
In vitro cultivation of tumor cells as multicellular tumor
spheroids (MTS) has greatly contributed to the understanding of
the role of the cellular micro-environment in tumor biology (for
review see \cite{Sutherland,Kunz}). These spherical cell
aggregates mimic avascular tumor stages or micro-metastases in
many aspects and have been studied intensively as an experimental
model reflecting an in vivo-like micro-milieu with 3D metabolic
gradients. With increasing size, most MCTS not only exhibit
proliferation gradients from the periphery towards the center but
they also develop a spheroid type-specific nutrient supply
pattern, such as radial oxygen partial pressure gradients.
Similarly, MCTS of a variety of tumor cell lines exhibit a
concentric histo-morphology, with a necrotic core surrounded by a
viable cell rim. The spherical symmetry is an important
prerequisite for investigating the effect of environmental factors
on cell proliferation and viability in a 3D environment on a
quantitative basis \cite{Walenta}.} \cite{Kunz}. They are used to
study how local micro-environments affect cellular growth/decay,
viability, and therapeutic response (see \cite{Sutherland}). MTS
are often combined with 3D medical imaging \cite{Hu}, 4D confocal imaging\footnote{%
Four--dimensional (4D) imaging of biological specimens (3D image
reconstruction of the same living sample at different time points), is an
application of confocal microscopy with fluorescence probes.} \cite%
{Khoshyomn}, 3D video holography through living tissue\footnote{%
Holographic coherence-domain imaging records full-frame depth resolved
images throughout living multicellular tumor spheroids in vitro, without
computed tomography, allowing real-time video fly-through under interactive
control of the operator.} \cite{Jeong} and 3D metabolic imaging imaging
bioluminescence\footnote{%
Imaging bioluminescence technique allows the mapping of metabolite
concentrations (e.g., ATP, glucose, and lactate) in cryosections
of spheroid sections at a high spatial resolution
\cite{Walenta2}.} \cite{Walenta}. MTS provide, allowing strictly
controlled nutritional and mechanical conditions, excellent
experimental patterns to test the validity of the proposed
mathematical models of tumor growth/decay \cite{Preziosi}.

A number of mathematical models of \emph{avascular tumor growth}
inside the MTS were reviewed in \cite{Roose}. These were generally
divided into continuum cell population models described by
diffusion partial differential equations (PDEs) of continuum
mechanics \cite{GaneshSprBig,GCompl} combined with chemical
kinetics, and discrete cell population models described by
ordinary differential equations (ODEs). Besides, in many cell
population models it is possible to empirically demonstrate the
presence of \emph{attractors} that operate starting from different
initial conditions \cite{TacaNODY,StrAttr}.

A general model of multi--phase tumor growth (inside the MTS) is
given in \cite{Roose} by the parabolic reaction--diffusion
PDE,\footnote{Reaction--diffusion systems are PDE--models that
describe how the concentration of one or more substances
distributed in space changes under the influence of two processes:
local (bio)chemical reactions in which the substances are
converted into each other, and diffusion which causes the
substances to spread out in space. They have the form of
semi-linear parabolic partial differential equations.}
\begin{equation}
\partial _{t}\Phi _{i}=\nabla \cdot (D_{i}\Phi _{i})-\nabla \cdot (\mathbf{v}%
_{i}\Phi _{i})+\lambda _{i}(\Phi _{i},C_{i})-\mu _{i}(\Phi _{i},C_{i})
\label{multiPh}
\end{equation}%
($\partial _{t}\equiv \partial /\partial {t}$), where for phase $i$, $\Phi
_{i}$ is the volume fraction ($\sum_{i}\Phi _{i}=1 $), $D_{i}$ is the random
motility or diffusion, $\lambda _{i}(\Phi _{i},C_{i})$ is the chemical and
phase dependent production, and $\mu _{i}(\Phi _{i},C_{i})$ is the chemical
and phase dependent degradation/death, and $\mathbf{v}_{i}$ is the cell
velocity defined by the constitutive equation
\begin{equation}
\mathbf{v}_{i}=-\mu \nabla p,  \label{constit}
\end{equation}%
where $\mu $ is a positive constant describing the viscous--like properties
of tumor cells and $p$ is the spheroid internal pressure.

In particular, the multi-phase equation (\ref{multiPh}) splits into two
heat--like mass--conservation PDEs \cite{Roose},
\begin{equation}
\partial _{t}\Phi ^{C}=S^{C}-\nabla \cdot (\mathbf{v}^{C}\Phi ^{C}),\qquad
\partial _{t}\Phi ^{F}=S^{F}-\nabla \cdot (\mathbf{v}^{F}\Phi ^{F}),
\label{2Ph}
\end{equation}%
where $\Phi ^{C}$ and $\Phi ^{F}$ are the tissue cell/matrix and fluid
volume fractions, respectively, $\mathbf{v}^{C}$ and $\mathbf{v}^{F}$ are\
the cell/matrix and the fluid velocities (both defined by their constitutive
equations of the form of (\ref{constit})), $S^{C}$ is the rate of production
of solid phase tumor tissue and $S^{F}$ is the creation/degradation of the
fluid phase. Conservation of matter in the tissue, $\Phi ^{C}+\Phi ^{F}=1$,
implies that $\nabla \cdot $($\mathbf{v}^{C}\Phi ^{C}$ +$\mathbf{v}^{F}\Phi
^{F}$) = $\Phi ^{C}+\Phi ^{F}$. The assumption that the tumor may be
described by two phases only implies that the new cell/matrix phase is
formed from the fluid phase and vice versa, so that $S^{C}+S^{F}=0$. The
detailed biochemistry of tumor growth can be coupled into the model above
through the growth term $S^{C}$, with equations added for nutrient
diffusion, see \cite{Roose} and references therein.

The multi-phase tumor growth model (\ref{multiPh}) has been derived from the
classical transport/mass conservation equations for different chemical
species inside the MTS \cite{Roose},
\begin{equation}
\partial _{t}C_{i}=P_{i}-\nabla \cdot \mathbf{N}_{i}.  \label{conEq}
\end{equation}%
Here $C_{i}$ are the concentrations of the chemical species, subindex $a$
for oxygen, $b$ for glucose, $c$ for lactate ion, $d$ for carbon dioxide, $e$
for bicarbonate ion, $f$ for chloride ion, and $g$ for hydrogen ion
concentration; $P_{i}$ is the net rate of consumption/production of the
chemical species both by tumor cells and due to the chemical reactions with
other species; and $\mathbf{N}_{i}$ is the flux of each of the chemical
species inside the tumor spheroid, given (in the simplest case of uncharged
molecules of glucose, $O_{2}$ and $CO_{2}$) by Fick's law,
\begin{equation*}
\mathbf{N}_{i}=-D_{i}\nabla C_{i},
\end{equation*}%
where $D_{i}$ are (positive) constant diffusion coefficients. In case of
charged molecules of ionic species, the flux $\mathbf{N}_{i}$ contains also
the (negative) gradient of the volume fractions $\Phi _{i}$.

In all above cases, tumor growth is, in terms of statistical mechanics, associated to \emph{entropy growth}. The more \emph{uncertainty} (measured as a number of tumor microstates) the tumor spheroid possesses, the larger is its entropy. Formally, we can apply \emph{Shannon formula} to the probability distributions $(p_1, \cdots, p_n)$ of cancer cells within the body,
$$S(p_1, \cdots, p_n) ~= ~- \sum_i p_i \log_2 p_i.$$ 
In other words, if we do not control the tumor growth, naturally it is governed by the Second Law of Thermodynamics:
$$\partial_{t}S~\geq ~0,$$
which is an expression of the universal law of increasing entropy, stating that the entropy (i.e., total number of tumor cells) of an isolated thermodynamic system (i.e., human body) which is not in equilibrium will tend to increase over time, approaching a maximum value at equilibrium (i.e., death threatening situation).

On the other hand, the \textit{Ricci flow equation} (or, the \textit{%
parabolic Einstein equation}), introduced by R. Hamilton in 1982
\cite{Ham82}, is the nonlinear heat--like evolution equation\footnote{%
The current hot topic in geometric topology is the Ricci flow, a
Riemannian evolution machinery that recently allowed G. Perelman
to prove the celebrated \textit{Poincar\'{e} Conjecture}, a
century--old mathematics problem (and one of the seven Millennium
Prize Problems of the Clay Mathematics Institute) -- and win him
the 2006 Fields Medal (which he declined in a public controversy)
\cite{Mackenzie}. The Poincar\'{e} Conjecture can roughly be put
as a question: Is a closed 3D manifold $M$ topologically a sphere
if every closed curve in $M$ can be shrunk continuously to a
point? In other words, Poincar\'{e} conjectured: A
simply-connected compact 3D manifold is diffeomorphic to the 3D
sphere $S^3$ (see e.g., \cite{Yau}).}
\begin{equation}
\partial _{t}g_{ij}=-2R_{ij},  \label{RF}
\end{equation}%
for a time--dependent Riemannian metric $g=g_{ij}(t)$ on a smooth real%
\footnote{%
For the related K\"{a}hler Ricci flow on complex manifolds, see e.g., \cite%
{GCompl,GaneshADG}} $n-$manifold $M$ with the Ricci curvature tensor $R_{ij}$%
.\footnote{This particular PDE (\ref{RF}) was chosen by Hamilton
for much the same reason that A. Einstein introduced the Ricci
tensor into his gravitation field equation,
$$R_{ij}-\frac{1}{2}g_{ij}R=8\pi T_{ij},$$ where $T_{ij}$ is the
energy--momentum tensor. Einstein needed a symmetric 2--index
tensor which arises naturally from the metric tensor $g_{ij}$ and
its first and second partial derivatives. The Ricci tensor
$R_{ij}$ is essentially the only possibility. In gravitation
theory and cosmology, the Ricci tensor has the volume--decreasing
effect (i.e., convergence of neighboring geodesics, see
\cite{HawkPenr}).} This equation roughly says that we can deform
any metric on a 2D surface or $n$D manifold by the negative of its
curvature; after \emph{normalization} (see Figure \ref{Ricci}),
the final state of such deformation will be a metric with constant
curvature. The factor of 2 in (\ref{RF}) is more or less
arbitrary, but the negative sign is essential to insure a kind of
complex \emph{volume exponential decay},\footnote{This complex
geometric process is globally similar to a generic exponential
decay ODE:
$$\dot{x}=-\lambda f(x),$$ for a positive function $f(x)$.
We can get some insight into its solution from the simple
exponential decay ODE,
$$\dot{x}=-\lambda x\qquad\text{with the solution}\qquad
x(t)=x_0{\rm e}^{-\lambda t},$$ (where $x=x(t)$ is the observed
quantity with its initial value $x_0$ and $\lambda$ is a positive
decay constant), as well as the corresponding $n$th order rate
equation (where $n>1$ is an integer),\[
\dot{x}=-\lambda x^{n}\qquad\text{with the solution}\qquad\frac{1}{x^{n-1}}=\frac{1}{{%
x_{0}}^{n-1}}+(n-1)\,\lambda t.
\]} since the Ricci flow equation (\ref{RF}) is a kind of
nonlinear generalization of the standard linear heat equation
\begin{equation}
\partial _{t}u=\Delta u.\label{h1}
\end{equation}
Like the heat equation (\ref{h1}), the Ricci flow equation
(\ref{RF}) is well behaved in forward time and acts as a kind of
smoothing operator (but is usually impossible to solve in backward
time). If some parts of a solid object are hot and others are
cold, then, under the heat equation, heat will flow from hot to
cold, so that the object gradually attains a uniform temperature.
To some extent the Ricci flow behaves similarly, so that the Ricci
curvature `tries' to become more uniform \cite{Milnor}, thus
depicting a monotonic \emph{entropy growth},\footnote{Note that
two different kinds of entropy functional have been introduced
into the theory of the Ricci flow, both motivated by concepts of
entropy in thermodynamics, statistical mechanics and information
theory. One is Hamilton's entropy, the other is Perelman's
entropy. While in Hamilton's entropy, the scalar curvature $R$ of
the metric $g_{ij}$ is viewed as the leading quantity of the
system and plays the role of a probability density, in Perelman's
entropy the leading quantity describing the system is the metric
$g_{ij}$ itself. Hamilton established the monotonicity of his
entropy along the volume--normalized Ricci flow on the 2--sphere
$S^2$ \cite{surface}. Perelman established the monotonicity of his
entropy along the Ricci flow in all dimensions \cite{Perel1}.}
$\partial_{t}S\geq 0$, which is due to the positive definiteness
of the metric $g_{ij}\geq 0$, and naturally implying the
\emph{arrow of time} \cite{Penr79,GaneshADG,GCompl}.

In a suitable local coordinate system, the Ricci flow equation (\ref{RF})
has a nonlinear heat--type form, as follows. At any time $t$, we can choose
local harmonic coordinates so that the coordinate functions are locally
defined harmonic functions in the metric $g(t)$. Then the Ricci flow takes
the form (see e.g., \cite{Anderson})
\begin{equation}
\partial _{t}g_{ij}=\Delta g_{ij}+Q_{ij}(g,\partial g),  \label{RH}
\end{equation}%
where $\Delta $ is the Laplace--Beltrami differential operator on functions
with respect to the metric $g$ and $Q$ is a lower--order term quadratic in $%
g $ and its first order partial derivatives. From the analysis of nonlinear
heat PDEs, one obtains existence and uniqueness of forward--time solutions
to the Ricci flow (\ref{RH}) on some time interval, starting at any smooth
initial metric $g_0=g_{ij}(0)$.

As a simple example of the Ricci flow equations (\ref{RF})--(\ref{RH}),
consider a round spherical boundary $S^2$ of the MTS of radius $r$. The
metric tensor on $S^2$ takes the form
\begin{equation*}
g_{ij}=r^{2}\hat{g}_{ij},
\end{equation*}%
where $\hat{g}_{ij}$ is the metric for a unit sphere, while the Ricci tensor
\begin{equation*}
R_{ij}=(n-1)\hat{g}_{ij}
\end{equation*}%
is independent of $r$. The Ricci flow equation on $S^2$ reduces to
\begin{equation*}
\dot{r}^{2}=-2(n-1),
\end{equation*}%
with solution
\begin{equation*}
r^{2}(t)=r^{2}(0)-2(n-1)t.
\end{equation*}%
Thus the boundary sphere $S^2$ collapses to a point in finite time (see \cite%
{Milnor}).

More generally, the following geometrization conjecture holds for an MTS
3--manifold $M$: Suppose that we start with a compact initial MTS--manifold $%
M_0$ whose Ricci tensor $R_{ij}$ is everywhere positive definite. Then, as $%
M_0$ shrinks to a point under the Ricci flow (\ref{RF}), it
becomes rounder and rounder. If we rescale the metric $g_{ij}$ on
$M_0$ so that the volume of $M_0$ remains constant, then $M_0$
converges towards another compact MTS--manifold $M_1$ of constant
positive curvature (see \cite{Ham82}).

In case of even more general MTS $3-$manifolds (outside the class
of positive Ricci curvature metrics), the situation is much more
complicated, as various singularities may arise. One way in which
singularities may arise during the Ricci flow is that a spherical
boundary $S^{2}=\partial M$ of an MTS $3-$manifold $M$ may
collapse to a point in finite time. Such collapses can be
eliminated by performing a kind of ``geometric surgery" on the MTS
manifold $M$, that is a sophisticated sequence of cutting and
pasting without accumulation of time errors\footnote{Hamilton's
idea was to perform surgery to cut off the singularities and
continue his flow after the surgery. If the flow develops
singularities again, one repeats the process of performing surgery
and continuing the flow. If one can prove there are only a finite
number of surgeries in any finite time interval, and if the
long-time behavior of solutions of the Ricci flow (\ref{RF}) with
surgery is well understood, then one would be able to recognize
the topological structure of the initial manifold. Thus Hamilton's
program, when carried out successfully, would lead to a proof of
the Poincar\'e conjecture and Thurston's geometrization conjecture
\cite{Yau}.} (see \cite{Perel2}). After a finite number of such
surgeries, each component either: (i) converges towards a
3--manifold of constant positive Ricci curvature which shrinks to
a point in finite time, or possibly (ii) converges towards an
$S^{2}\times S^{1}$ which shrinks to a circle $S^{1}$ in finite
time, or (iii) admits a ``thin--thick" decomposition of
\cite{Thurston}. Therefore, one can choose the surgery parameters
so that there is a well defined Ricci-flow-with surgery, that
exists for all time \cite{Perel2}.

In this paper we use the evolving geometric machinery of the
volume--decaying and entropy--growing Ricci flow $g(t)$, given by
equations (\ref{RF})--(\ref{RH}), for modelling general 3D
avascular MTS decay, corresponding to parabolic multi--phase
reaction--diffusion PDEs (\ref{multiPh})--(\ref{conEq}).

\section{Ricci flow and multi-phase avascular MTS decay control}

\subsection{Geometrization Conjecture}

Recall that geometry and topology of smooth surfaces are related by the
\textit{Gauss--Bonnet formula} for a closed surface $\Sigma $ (see, e.g.,
\cite{GaneshSprBig,GaneshADG})
\begin{equation}
\frac{1}{2\pi}\iint_{\Sigma }K\,dA= \chi (\Sigma )=2-2\,{\rm
gen}(\Sigma ), \label{GB}
\end{equation}
where $dA$ is the area element of a metric $g$ on $\Sigma $, $K$
is the Gaussian curvature, $\chi (\Sigma )$ is the Euler
characteristic of $\Sigma $ and ${\rm gen}(\Sigma )$ is its
\emph{genus}, or number of handles, of $\Sigma $. Every closed
surface $\Sigma $ admits a metric of constant Gaussian curvature
$K=+1,\,0$, or $-1$ and so is uniformized by elliptic, Euclidean,
or hyperbolic geometry, which respectively have ${\rm gen}(S^2)=0$
(sphere), ${\rm gen}(T^2)=1$ (torus) and ${\rm gen}(\Sigma )>1$
(torus with several holes). The integral (\ref{GB}) is a
\emph{topological invariant} of the surface $\Sigma $, always
equal to 2 for all topological spheres $S^2$ (that is, for all
closed surfaces without holes that can be continuously deformed
from the geometrical sphere) and always equal to 0 for the
topological torus $T^2$ (i.e., for all closed surfaces with one
hole or handle).

The general topological framework for the Ricci flow (\ref{RF}) is
Thurston's \textit{Geometrization Conjecture} \cite{Thurston}, which states
that the interior of any compact 3--manifold can be split in an essentially
unique way by disjoint embedded 2D spheres $S^{2}$ and tori $T^{2}$ into
pieces and each piece admits one of 8 geometric structures (including (i)
the 3D sphere $S^{3}$ with constant curvature $+1$; (ii) the 3D Euclidean
space $\mathbb{R}^{3}$ with constant curvature 0 and (iii) the 3D hyperbolic
space $\mathbb{H}^{3}$ with constant curvature $-1$).\footnote{%
Another five allowed geometric structures are represented by the following
examples: (iv) the product $S^{2}\times S^{1}$; (v) the product $\mathbb{H}%
^{2}\times S^{1}$ of hyperbolic plane and circle; (vi) a left invariant
Riemannian metric on the special linear group $SL(2,\mathbb{R})$; (vii) a
left invariant Riemannian metric on the solvable Poincar\'{e}-Lorentz group $%
E(1,1)$, which consists of rigid motions of a ($1+1)-$dimensional space-time
provided with the flat metric $dt^{2}-dx^{2}$; (viii) a left invariant
metric on the nilpotent Heisenberg group, consisting of $3\times 3$ matrices
of the form
\par
$\left[
\begin{array}{ccc}
1 & \ast & \ast \\
0 & 1 & \ast \\
0 & 0 & 1%
\end{array}%
\right] .$ In each case, the universal covering of the indicated manifold
provides a canonical model for the corresponding geometry \cite{Milnor}.}
The geometrization conjecture (which has the Poincar\'{e} Conjecture as a
special case) would give us a link between the geometry and topology of MTS
3--manifolds, analogous in spirit to the case of 2D surfaces.

In higher dimensions, the Gaussian curvature $K$ corresponds to the Riemann
curvature tensor $\mathfrak{Rm}$ on a smooth $n-$manifold $M$, which is in
local coordinates on $M$ denoted by its $(4,0)-$components $R_{ijkl}$, or
its $(3,1)-$components $R_{ijk}^{l}$ (see Appendix, as well as e.g., \cite%
{GaneshSprBig,GaneshADG}). The trace (or, contraction) of $\mathfrak{Rm}$,
using the inverse metric tensor $g^{ij}=(g_{ij})^{-1}$, is the Ricci tensor $%
\mathfrak{Rc}$, the 3D curvature tensor, which is in a local coordinate
system $\{x^{i}\}_{i=1}^{n}$ defined in an open set $U\subset M$, given by
\begin{equation*}
R_{ij}=\mathrm{tr}(\mathfrak{Rm})=g^{kl}R_{ijkl}
\end{equation*}%
(using Einstein's summation convention), while the \textit{scalar curvature}
is now given by the second contraction of $\mathfrak{Rm}$ as
\begin{equation*}
R=\mathrm{tr}(\mathfrak{Rc})=g^{ij}R_{ij}.
\end{equation*}

In general, the Ricci flow $g_{ij}(t)$ is a one--parameter family of
Riemannian metrics on a compact $n-$manifold $M$ governed by the equation (%
\ref{RF}), which has a unique solution for a short time for an arbitrary
smooth metric $g_{ij}$ on $M$ \cite{Ham82}. If $\mathfrak{Rc}>0$ at any
local point $x=\{x^i\}$ on $M$, then the Ricci flow (\ref{RF}) contracts the
metric $g_{ij}(t)$ near $x$, to the future, while if $\mathfrak{Rc}<0$, then
the flow (\ref{RF}) expands $g_{ij}(t)$ near $x$. The solution metric $%
g_{ij}(t)$ of the Ricci flow equation (\ref{RF}) shrinks in positive Ricci
curvature direction while it expands in the negative Ricci curvature
direction, because of the minus sign in the front of the Ricci tensor $%
R_{ij} $. In particular, in 2D, on a sphere $S^{2}$, any metric of
positive Gaussian curvature will shrink to a point in finite time.
At a general point, there will be directions of positive and
negative Ricci curvature along which the metric will locally contract or expand (see \cite{Anderson}%
). In 3D, if a simply-connected compact 3--manifold $M$ has a
Riemannian metric $g_{ij}$ with positive Ricci curvature then it
is diffeomorphic to the 3--sphere $S^3$ \cite{Ham82}.

All three Riemannian curvatures ($R,\mathfrak{Rc}$ and
$\mathfrak{Rm}$),
as well as the associated volume forms, \emph{evolve} during the Ricci flow (%
\ref{RF}).

\subsection{MTS evolution under the Ricci flow}

The Ricci flow evolution equation (\ref{RF}) for the metric tensor $g_{ij}$
implies the evolution equation for the Riemann curvature tensor $\mathfrak{Rm%
}$,
\begin{equation}
\partial _{t}\mathfrak{Rm}=\triangle \mathfrak{Rm}+Q_n,  \label{RMflow}
\end{equation}%
where $Q_n$ is a certain quadratic expression of the Riemann curvatures.
From the general $n-$curvature expression (\ref{RMflow}) we have two special
cases important for MTS--evolution:\footnote{%
By expanding the maximum principle for tensors, Hamilton proved that Ricci
flow $g(t)$ given by (\ref{RF}) preserves the positivity of the Ricci tensor
$\mathfrak{Rc}$ in 3D (as well as of the Riemann curvature tensor $\mathfrak{%
Rm}$ in all dimensions); moreover, the eigenvalues of the Ricci
tensor in 3D (and of the curvature operator $\mathfrak{Rm}$ in 4D)
are getting pinched point-wisely as the curvature is getting large
\cite{Ham82,4-manifold}. This observation allowed him to prove the
convergence results: the evolving metrics (on a compact manifold)
of positive Ricci curvature in 3D (or positive Riemann curvature
in 4D) converge, modulo scaling, to metrics of constant positive
curvature.
\par
However, without assumptions on curvature, the long time behavior of the
metric evolving by Ricci flow may be more complicated \cite{Perel1}. In
particular, as $t$ approaches some finite time $T$, the curvatures may
become arbitrarily large in some region while staying bounded in its
complement. On the other hand, Hamilton \cite{Harnack} discovered a
remarkable property of solutions with nonnegative curvature tensor $%
\mathfrak{Rm}$ in arbitrary dimension, called the \emph{differential Harnack
inequality}, which allows, in particular, to compare the curvatures of the
solution of (\ref{RF}) at different points and different times.}

\begin{itemize}
\item The 3D evolution equation for the Ricci curvature tensor $\mathfrak{Rc}
$ on an MTS 3--manifold $M$,
\begin{equation}
\partial _{t}\mathfrak{Rc}=\triangle \mathfrak{Rc}+Q_3,  \label{Rcflow}
\end{equation}%
where $Q_3$ is a certain quadratic expression of the Ricci curvatures; and

\item The 2D evolution equation for the scalar surface curvature $R$,
\begin{equation}
\partial _{t}R=\triangle R+2|\mathfrak{Rc}|^{2},  \label{DTR}
\end{equation}%
which holds both on an MTS 3--manifold $M$ and on its 2D boundary surface $%
\partial M$. Therefore, by the \textit{maximum principle}, the minimum of $R$
is non--decreasing along the flow $g(t)$, both on $M$ and on $\partial M$
(see \cite{Perel1}).
\end{itemize}

Let us now see in detail how various MTS--related geometric quantities
evolve given the short-time solution of the Ricci flow equation (\ref{RF})
on an MTS 3--manifold $M$. Let us first calculate the \textit{variation
formulas} for the Christoffel symbols and curvature tensors on $M$ and then
the corresponding evolution equations (see \cite{Ham82,CaoChow,ChowKnopf}).
If $g(s)$ is a one--parameter family of metrics on $M$ with
\begin{equation*}
\partial _{s}g_{ij}=v_{ij},
\end{equation*}%
then the variation of the Christoffel symbols $\Gamma _{ij}^{k}$ on $M$ is
given by
\begin{equation}
\partial _{s}\Gamma _{ij}^{k}=\frac{1}{2}g^{kl}\left( \nabla
_{i}v_{jl}+\nabla _{j}v_{il}-\nabla _{l}v_{ij}\right) ,  \label{VC}
\end{equation}%
from which follows the evolution of the Christoffel symbols $\Gamma
_{ij}^{k} $ under the Ricci flow $g(t)$ on $M$ given by (\ref{RF}),%
\begin{equation*}
\partial _{t}\Gamma _{ij}^{k}=-g^{kl}\left( \nabla _{i}R_{jl}+\nabla
_{j}R_{il}-\nabla _{l}R_{ij}\right) .
\end{equation*}

From (\ref{VC}) we calculate the variation of the Ricci tensor $R_{ij}$ on $%
M $ as
\begin{equation}
\partial _{s}R_{ij}=\nabla _{m}\left( \partial _{s}\Gamma _{ij}^{m}\right)
-\nabla _{i}\left( \partial _{s}\Gamma _{mj}^{m}\right) ,  \label{VRic}
\end{equation}%
and the variation of scalar curvature $R$ on $M$ by
\begin{equation}
\partial _{s}R=-\Delta V+\mathrm{div}(\mathrm{div\,}v)-\left\langle v,%
\mathfrak{Rc}\right\rangle ,  \label{VR}
\end{equation}%
where $V=g^{ij}v_{ij}=\mathrm{tr}(v)$ is the trace of $v=(v_{ij})$.

If an MTS $3-$manifold $M$ is oriented, then the \textit{volume} $3-$form on
$M$ is given, in a positively oriented local coordinate system $\{x^{i}\}\in
U\subset M$, by\footnote{%
Extension to higher--dimensional Riemannian manifolds is obvious \cite%
{GaneshADG}; also, for related volume forms on symplectic manifolds, see
\cite{VladSiam}}
\begin{equation}
d\mu =\sqrt{\det (g_{ij})}\,dx^{1}\wedge dx^{2}\wedge dx^{3}.  \label{dmu}
\end{equation}%
If $\partial _{s}g_{ij}=v_{ij},$ then
\begin{equation*}
\partial _{s}d\mu =\frac{1}{2}Vd\mu .
\end{equation*}%
The evolution of the volume form $d\mu $ under the Ricci flow
$g(t)$ on $M$ is given by the exponential decay/growth relation
with the scalar curvature $R$ as the rate constant,
\begin{equation}
\partial _{t}d\mu =-Rd\mu, \label{dtmu}
\end{equation}
which gives an exponential decay for $R>0$ (elliptic geometry) and
exponential growth for $R<0$ (hyperbolic geometry). The elementary
volume evolution (\ref{dtmu}) implies the integral form of the
exponential relation for the total MTS--volume
\begin{equation*}
\mathrm{vol}(g)=\int_{M}d\mu,
\end{equation*}%
in the form
\begin{equation*}
\partial _{t}\mathrm{vol}(g(t))=-\int_{M}Rd\mu,
\end{equation*}
which again gives an exponential decay for elliptic $R>0$ and
exponential growth for hyperbolic $R<0$.

This is a crucial point for the tumor decay control: we need to
keep the elliptic geometry of the MTS -- by all possible means.
And naturally -- it will be so, because it started as a spherical
shape with $R>0$. We just need to keep the MTS in this shape and
prevent any hyperbolic distortions of $R<0$. The, it will
naturally have an exponential decay.

\begin{figure}[h]
\centerline{\includegraphics[width=6cm]{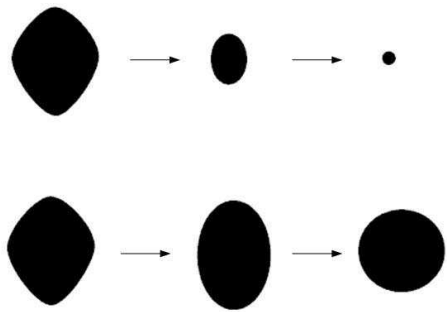}} \caption{An
example of Ricci flow normalization: unnormalized flow (up) and
normalized flow (down).} \label{Ricci}
\end{figure}

On the other hand, if we are not able to keep the positivity of
the scalar curvature of the MTS, and thus prevent its expanding to
in infinity (`deadly' hyperbolic case), we can also consider the
\textit{normalized Ricci flow} of the MTS on its 3--manifold $M$
(see Figure \ref{Ricci} as well as ref. \cite{CaoChow}):
\begin{equation}
\partial _{t}\hat{g}_{ij}=-2\hat{R}_{ij}+\frac{2}{n}\hat{r}\hat{g}_{ij},
\label{NRF}
\end{equation}%
where
\begin{equation*}
\hat{r}=\mathrm{vol}(\hat{g})^{-1}\int_{M}\hat{R}d\mu
\end{equation*}%
is the average scalar curvature on $M$. We then have the MTS
\emph{volume conservation law:}
\begin{equation*}
\partial _{t}\mathrm{vol}(\hat{g}(t))=0.
\end{equation*}

To study the long--time existence of the normalized Ricci flow (\ref{NRF})
on an MTS $3-$mani-fold $M$, it is important to know what kind of curvature
conditions are preserved under the equation. In general, the Ricci flow $%
g(t) $ on $M$ tends to preserve some kind of positivity of curvatures. For
example, positive scalar curvature $R$ is preserved both on $M$ and on its
boundary $\partial M$. This follows from applying the maximum principle to
the evolution equation (\ref{DTR}) for scalar curvature $R$ both on $M$ and
on $\partial M$. Also, positive Ricci curvature is preserved under the Ricci
flow on $M$. (This is a special feature of 3D and is related to the fact
that the Riemann curvature tensor may be recovered algebraically from the
Ricci tensor and the metric in 3D \cite{CaoChow}.)

In particular, we have the following result (see \cite{surface}) for
MTS--surfaces $\partial M$: Let $\partial M$ be a closed MTS--surface. Then
for any initial 2D metric $g_{0}$ on $\partial M$, the solution to the
normalized Ricci flow (\ref{NRF}) on $\partial M$ exists for all time.
Moreover, (i) If the Euler characteristic of $\partial M$ is non--positive,
then the solution metric $g(t)$ on $\partial M$ converges to a constant
curvature metric as $t\rightarrow \infty $; and (ii) If the scalar curvature
$R$ of the initial metric $g_{0}$ is positive, then the solution metric $%
g(t) $ on $\partial M$ converges to a positive constant curvature metric as $%
t\rightarrow \infty .$ (For surfaces with non--positive Euler
characteristic, the proof is based primarily on maximum principle estimates
for the scalar curvature.)

In other words, the normalized Ricci flow of the MTS will make it
completely round with a geometrical sphere shell -- ideal for
surgical removal. This is our second option for the MTS control.
If we cannot force it to exponential decay, then we must try to
normalize into a round spherical shell -- which is suitable for
surgical removal.

The negative flow of the total MTS--volume $\mathrm{vol}(g(t)) $ is the \textit{Einstein--Hilbert functional}, given by (see \cite%
{MTW,CaoChow,Anderson})
\begin{equation*}
E(g)=\int_{M}Rd\mu =-\partial _{t}\mathrm{vol}(g(t)).
\end{equation*}%
If we put $\partial _{s}g_{ij}=v_{ij},$ we have%
\begin{eqnarray*}
\partial _{s}E(g) &=&\int_{M}\left( -\Delta V+\mathrm{div}(\mathrm{div\,}%
v)-\left\langle v,\mathfrak{Rc}\right\rangle +\frac{1}{2}RV\right) d\mu \\
&=&\int_{M}\left\langle v,\frac{1}{2}Rg_{ij}-R_{ij}\right\rangle d\mu ,
\end{eqnarray*}%
so the critical points of $E(g)$ satisfy \emph{Einstein's
equation}
\begin{equation*}
\frac{1}{2}Rg_{ij}-R_{ij}=0.
\end{equation*}%
The gradient flow of $E(g)$ on $M$, given by%
\begin{equation*}
\partial _{t}g_{ij}=2\left( \nabla E(g)\right) _{ij}=Rg_{ij}-2R_{ij},
\end{equation*}%
is almost the Ricci flow (\ref{RF}). Thus, Einstein metrics are
the fixed points of the Ricci flow $g(t)$ on $M$.\footnote{In 3D
manifolds, Einstein metrics are metrics with constant curvature.
However, along the way, the deformation will encounter
singularities. The major question, resolved by Perelman, was how
to find a way to describe all possible singularities.}

Let $\Delta $ denote the Laplacian acting on functions on an MTS $3-$%
mani-fold $M$, which is in local coordinates $\{x^{i}\}\in U\subset M$ given
by%
\begin{equation*}
\Delta =g^{ij}\nabla _{i}\nabla _{j}=g^{ij}\left( \partial _{ij}-\Gamma
_{ij}^{k}\partial _{k}\right) .
\end{equation*}%
For any smooth function $f$ on $M$ we have \cite{Ham82,ChowKnopf}%
\begin{eqnarray*}
\Delta \nabla _{i}f &=&\nabla _{i}\Delta f+R_{ij}\nabla _{j}f,\qquad \text{%
and} \\
\Delta |\nabla f|^{2} &=&2|\nabla _{i}\nabla _{j}f|^{2}+2R_{ij}\nabla
_{i}f\nabla _{j}f+2\nabla _{i}f\nabla _{i}\Delta f.
\end{eqnarray*}%
From this it follows that if we have
\begin{eqnarray*}
\mathfrak{Rc} &\geq &0,\qquad \Delta f\equiv 0,\qquad |\nabla f|\equiv
1,\qquad \text{then} \\
\nabla \nabla f &\equiv &0\qquad \text{and\qquad }\mathfrak{Rc}(\nabla
f,\nabla f)\equiv 0.
\end{eqnarray*}

Using $\Delta ,$ we can write the linear heat equation on $M$ as
\begin{equation*}
\partial _{t}u=\Delta u,
\end{equation*}%
where $u$ is the MTS--temperature. In particular, the Laplacian acting on
functions with respect to $g(t)$ will be denoted by $\Delta _{g(t)}$. If $%
(M,g(t))$ is a solution to the Ricci flow equation (\ref{RF}), then we have%
\begin{equation*}
\partial _{t}\Delta _{g(t)}=2R_{ij}\nabla _{i}\nabla _{j}.
\end{equation*}

Now, the evolution equation (\ref{DTR}) for the scalar curvature $R$ under
the Ricci flow (\ref{RF}) follows from (\ref{VR}). Using equation (\ref{cB})
from Appendix, we have:
\begin{equation*}
\mathrm{div}(\mathfrak{Rc})=\frac{1}{2}\nabla R,\quad\text{so that}\quad
\mathrm{div}(\mathrm{div}(\mathfrak{Rc})) = \frac{1}{2}\Delta R,
\end{equation*}
showing again that the scalar curvature $R$ satisfies a heat--type equation
with a quadratic nonlinearity both on an MTS 3--manifold $M$ and on its 2D
boundary surface $\partial M$.

Next we will find the exact form of the evolution equation (\ref{Rcflow})
for the Ricci tensor $\mathfrak{Rc}$ under the Ricci flow $g(t)$ given by (%
\ref{RF}) on an MTS 3--manifold $M$. (Note that in higher dimensions, the
appropriate formula would involve the whole Riemann curvature tensor $%
\mathfrak{Rm}$.) In general, given a variation $\partial _{s}g_{ij}=v_{ij}$,
from (\ref{VRic}) we get
\begin{equation*}
\partial _{s}R_{ij}=\frac{1}{2}\left( \Delta _{L}v_{ij}+\nabla _{i}\nabla
_{j}V-\nabla _{i}(\mathrm{div\,}v)_{j}-\nabla _{j}(\mathrm{div\,}%
v)_{i}\right) ,
\end{equation*}%
where $\Delta _{L}$ denotes the so--called Lichnerowicz Laplacian (which
depends on $\mathfrak{Rm})$ (see \cite{Ham82,ChowKnopf}). Since%
\begin{equation*}
\nabla _{i}\nabla _{j}R-\nabla _{i}(\mathrm{div}(\mathfrak{Rc}))_{j}-\nabla
_{j}(\mathrm{div}(\mathfrak{Rc}))_{i}=0,
\end{equation*}%
by (\ref{cB}) (after some algebra) we get that under the Ricci flow (\ref{RF}%
) the evolution equation for the Ricci tensor $\mathfrak{Rc}$ on $M$ is
\begin{equation*}
\partial _{t}R_{ij}=\Delta R_{ij}+3RR_{ij}-6R_{im}R_{jm}+\left( 2|\mathfrak{%
Rc}|^{2}-R^{2}\right) g_{ij}.
\end{equation*}%
So, just as in case of the evolution (\ref{DTR}) of the scalar curvature $%
\partial _{t}R$ (both on $M$ and on its boundary $\partial M$), we get a
heat--type evolution equation with a quadratic nonlinearity for $\partial
_{t}R_{ij}$, which means that positive Ricci curvature ($\mathfrak{Rc}>0$)
of elliptic MTS--geometry is preserved under the Ricci flow $g(t)$ on $M$.

More generally, we have the following result for MTS 3--manifolds (see \cite%
{Ham82}): Let $(M,g_{0})$ be a compact Riemannian MTS $3-$manifold
with positive Ricci curvature $\mathfrak{Rc}$. Then there exists a
unique solution to the normalized Ricci flow $g(t)$ on $M$ with
$g(0)=g_{0}$ for all time and the metrics $g(t)$ converge
exponentially fast to a constant positive sectional curvature
metric $g_{\infty }$ on $M$. In particular, $M$
is diffeomorphic to a 3D sphere $S^3$. (As a consequence, such an MST $3-$%
manifold $M$ is necessarily diffeomorphic to a quotient of the
$3-$sphere by a finite group of isometries. It follows that given
any homotopy $3-$sphere, if one can show that it admits a metric
with positive Ricci curvature, then the Poincar\'{e} Conjecture
would follow \cite{CaoChow}.) In particular, compact and closed
$3-$manifolds which admit a non-singular solution can also be
decomposed into geometric pieces \cite{non-singular}.

\subsection{Ricci breathers and solitons}

Recall that \emph{breathers} are solitonic structures given by localized
periodic solutions of some nonlinear soliton PDEs, including the exactly
solvable sine-Gordon equation\footnote{%
An exact solution $u=u(x,t)$ of the (1+1)D sine--Gordon equation
\begin{equation*}
\frac{\partial^2 u}{\partial t^2} = \frac{\partial^2 u}{\partial x^2} - \sin
u,
\end{equation*}
is \cite{Ablowitz}
\begin{equation*}
u = 4 \arctan\left(\frac{\sqrt{1-\omega^2}\;\cos(\omega t)}{\omega\;\cosh(%
\sqrt{1-\omega^2}\; x)}\right),
\end{equation*}
which, for $\omega< 1$, is periodic in time $t$ and decays exponentially
when moving away from $x = 0$.} and the focusing nonlinear Schr\"odinger
equation.\footnote{%
The focusing nonlinear Schr\"odinger equation is the dispersive
complex-valued (1+1)D PDE \cite{Akhmediev},
\begin{equation*}
i\,\frac{\partial u}{\partial t} + \frac{\partial^2 u}{\partial x^2} + |u|^2
u = 0,\qquad(i=\sqrt{-1},\,u=u(x,t))
\end{equation*}
with a breather solution of the form:
\begin{equation*}
u = \left( \frac{2\, b^2 \cosh(\theta) + 2\, i\, b\, \sqrt{2-b^2}\;
\sinh(\theta)} {2\, \cosh(\theta)-\sqrt{2}\,\sqrt{2-b^2} \cos(a\, b\, x)} -
1 \right)\; a\; \exp(i\, a^2\, t) \quad\text{with}\quad \theta=a^2\,b\,\sqrt{%
2-b^2}\;t,
\end{equation*}
which gives breathers periodic in space $x$ and approaching the
uniform value $a$ when moving away from the focus time $t = 0$.}

A metric $g_{ij}(t)$ evolving by the Ricci flow $g(t)$ given by
(\ref{RF})
on an MTS 3--manifold $M$ is called a \emph{Ricci breather}, if for some $%
t_{1}<t_{2}$ and $\alpha >0$ the metrics $\alpha g_{ij}(t_{1})$ and $%
g_{ij}(t_{2})$ differ only by a diffeomorphism; the cases $\alpha =1,\alpha
<1,\alpha >1$ correspond to steady, shrinking and expanding breathers,
respectively. Trivial breathers on $M$, for which the metrics $g_{ij}(t_{1})$
and $g_{ij}(t_{2})$ differ only by diffeomorphism and scaling for each pair
of $t_{1}$ and $t_{2}$, are called \textit{Ricci solitons}. Thus, if one
considers Ricci flow as a dynamical system on the space of Riemannian
metrics modulo diffeomorphism and scaling, then breathers and solitons
correspond to periodic orbits and fixed points respectively. At each time
the Ricci soliton metric satisfies on $M$ an equation of the form \cite%
{Perel1}
\begin{equation*}
R_{ij}+cg_{ij}+\nabla _{i}b_{j}+\nabla _{j}b_{i}=0,
\end{equation*}%
where $c$ is a number and $b_{i}$ is a one-form; in particular, when $b_{i}=%
\frac{1}{2}\nabla _{i}a$ for some function $a$ on $M,$ we get a gradient
Ricci soliton. An important example of a gradient shrinking soliton is the
\textit{Gaussian soliton}, for which the metric $g_{ij}$ is just the
Euclidean metric on $\mathbb{R}^{3}$, $c=1$ and $a=-|x|^{2}/2$.

\subsection{Heat equation and Ricci entropy}

Given a $C^{2}$ function $u:M\rightarrow \mathbb{R}$ on a
Riemannian MTS $3-$manifold $M$, its Laplacian is defined in local
coordinates $\left\{ x^{i}\right\}\in U\subset M $ to be
\begin{equation*}
\Delta u=\text{\textrm{tr}}_{g}\left( \nabla ^{2}u\right)
=g^{ij}\nabla _{i}\nabla _{j}u,
\end{equation*}%
where $\nabla _{i}$ is its associated covariant derivative
(Levi--Civita connection, see Appendix). We say that a $C^{2}$
function $u:M\times \lbrack 0,T)\rightarrow \mathbb{R},$ where
$T\in (0,\infty ],$ is a solution to the heat equation on $M$ if
\begin{equation}
\partial _{t}u=\Delta u.  \label{heat1}
\end{equation}%
One of the most important properties satisfied by the heat
equation is the \textit{maximum principle}, which says that for
any smooth solution to the heat equation, whatever point-wise
bounds hold at $t=0$ also hold for $t>0$ \cite{CaoChow}. More
precisely, we can state: Let $u:M\times \lbrack
0,T)\rightarrow \mathbb{R}$ be a $C^{2}$ solution to the heat equation (\ref%
{heat1})\ on a complete Riemannian MTS $3-$manifold $M$. If
$C_{1}\leq u\left( x,0\right) \leq C_{2}$ for all $x\in M,$ for
some constants $C_{1},C_{2}\in \mathbb{R},$ then $C_{1}\leq
u\left( x,t\right) \leq C_{2}$ for all $x\in M$ and $t\in \lbrack
0,T).$ This property exhibits the smoothing behavior of the heat
equation (\ref{heat1}) on $M$.

Now, consider Perelman's \emph{entropy functional} \cite{Perel1} on an MTS
3--manifold $M$
\begin{equation}
\mathcal{F}=\int_{M}(R+|\nabla f|^{2}){\mathrm{e}}^{-f}d\mu  \label{F}
\end{equation}%
for a Riemannian metric $g_{ij}$ and a (temperature-like) scalar function $f$
on a closed 3--manifold $M$, where $d\mu $ is the volume 3--form (\ref{dmu}%
). During the Ricci flow (\ref{RF}), $\mathcal{F}$ evolves on $M$ as
\begin{equation}
\partial _{t}\mathcal{F}=2\int |R_{ij}+\nabla _{i}\nabla _{j}f|^{2}{\mathrm{e%
}^{-f}}d\mu {.}  \label{dF}
\end{equation}%
Now, define $\lambda (g_{ij})=\inf \mathcal{F}(g_{ij},f),$ where infimum is
taken over all smooth $f,$ satisfying
\begin{equation}
\int_{M}{\mathrm{e}^{-f}}d\mu =1.  \label{eDm}
\end{equation}%
$\lambda (g_{ij})$ is the lowest eigenvalue of the operator
$-4\triangle +R.$ Then the entropy evolution formula (\ref{dF})
implies that $\lambda (g_{ij}(t))$ is nondecreasing in $t,$ and
moreover, if $\lambda
(t_{1})=\lambda (t_{2}),$ then for $t\in \lbrack t_{1},t_{2}]$ we have $%
R_{ij}+\nabla _{i}\nabla _{j}f=0$ for $f$ which minimizes
$\mathcal{F}$ on $M $ \cite{Perel1}. Thus a steady breather on $M$
is necessarily a steady soliton.

If we define the conjugate \emph{heat operator} on $M$ as
\begin{equation*}
\Box ^{\ast }=-\partial /\partial t-\triangle +R
\end{equation*}%
then we have the \emph{conjugate heat equation}\footnote{In
\cite{Perel1} Perelman stated a differential Li--Yau--Hamilton
(LYH) type inequality \cite{Hsu2} for the fundamental solution
$u=u(x,t)$ of the conjugate heat equation (\ref{conHeat1}) on a
closed $n-$manifold $M$ evolving by the Ricci flow (\ref{RF}). Let
$p\in M$ and
\begin{equation*}
u=(4\pi \tau )^{-\frac{n}{2}}\mathrm{e}^{-f}
\end{equation*}%
be the fundamental solution of the conjugate heat equation in
$M\times (0,T)$,
\begin{equation*}
\Box ^{\ast }u=0,\qquad\text{or}\qquad\partial _{t}u+\Delta u=Ru,
\end{equation*}%
where $\tau =T-t$ and $R=R(\cdot ,t)$ is the scalar curvature of
$M$ with respect to the metric $g(t)$ with $\lim_{t\nearrow
T}u=\delta _{p}$ (in the distribution sense), where $\delta _{p}$
is the delta--mass at $p$. Let
\begin{equation*}
v=[\tau (2\Delta f-|\nabla f|^{2}+R)+f-n]u,
\end{equation*}%
where $\tau =T-t$. Then we have a differential LYH--type
inequality
\begin{equation}
v(x,t)\leq 0\quad \text{ in \ }M\times (0,T).  \label{ineq1}
\end{equation}%
This result was used by Perelman to give a proof of the \textit{%
pseudolocality theorem} \cite{Perel1} which roughly said that
almost Euclidean regions of large curvature in closed manifold
with metric evolving by Ricci flow $g(t)$ given by (\ref{RF})
remain localized.
\par
In particular, let $(M,g(t))$, $0\leq t\leq T$, $\partial M\neq \phi $, be a compact $3-$%
manifold (like MTS) with metric $g(t)$ evolving by the Ricci flow $g(t)$ given by (\ref%
{RF}) such that the second fundamental form of the surface
$\partial M$ with respect to the unit outward normal $\partial
/\partial \nu $ of $\partial M$ is uniformly bounded below on
$\partial M\times \lbrack 0,T]$. A global Li--Yau gradient
estimate \cite{LY}\ for the solution of the generalized conjugate
heat equation was proved in \cite{Hsu2} (using a a variation of
the method of P.~Li and S.T.~Yau, \cite{LY}) on such a manifold
with Neumann boundary condition.} \cite{Perel1}
\begin{equation}
\Box ^{\ast }u=0. \label{conHeat1}
\end{equation}

The entropy functional (\ref{F}) is nondecreasing under the
following coupled \emph{Ricci--heat flow} on $M$ \cite{Li07}
\begin{eqnarray}
\partial _{t}g_{ij} &=&-2R_{ij},  \notag \\
\partial _{t}u &=&-\Delta u-\frac{|\nabla u|^{2}}{u}+\frac{R}{2}u,
\label{conHeat}
\end{eqnarray}%
where the modified conjugate heat equation (\ref{conHeat}) ensures
\begin{equation*}
\int_{M}u^{2}d\mu =1
\end{equation*}%
to be preserved by the Ricci flow $g(t)$ on $M$. If we define $\ u=\mathrm{e}%
^{-\frac{f}{2}}$, then (\ref{conHeat}) is equivalent to $f-$evolution
equation on $M$,
\begin{equation*}
\partial _{t}f=-\Delta f+|\nabla f|^{2}-R,
\end{equation*}%
which instead preserves (\ref{eDm}).

\subsection{Thermodynamic analogy}

Perelman's functional $\mathcal{F}$ is analogous to negative entropy \cite%
{Perel1}. Recall that thermodynamic \textit{partition function}
for a generic canonical ensemble at temperature $\beta ^{-1}$ is
given by
\begin{equation}
Z=\int \mathrm{e}^{{-\beta E}}d\omega (E),  \label{Z}
\end{equation}%
where $\omega (E)$ is a `density measure', which does not depend
on $\beta .$ From it, the \textit{average energy} is given by
\[
\left\langle E\right\rangle =-\partial _{\beta }\ln Z,
\]%
the \textit{entropy }is
\[
S=\beta \left\langle E\right\rangle +\ln Z,
\]%
and the \textit{fluctuation }is
\[
\sigma =\left\langle (E-\left\langle E\right\rangle
)^{2}\right\rangle =\partial _{\beta ^{2}}\ln Z.
\]

If we now fix a closed MTS $3-$manifold $M$ with a probability
measure $m$ and a metric $g_{ij}(\tau )$ that depends on the
temperature $\tau $, then according to equation
\[
\partial _{\tau }g_{ij}=2(R_{ij}+\nabla _{i}\nabla _{j}f),
\]%
the partition function (\ref{Z}) is given by
\begin{equation}
\ln Z=\int (-f+\frac{n}{2})dm.  \label{lnZ}
\end{equation}%
From (\ref{lnZ}) we get (see \cite{Perel1})
\begin{eqnarray*}
\left\langle E\right\rangle  &=&-\tau ^{2}\int_{M}(R+|\nabla f|^{2}-\frac{n}{%
2\tau })dm, \\
S &=&-\int_{M}(\tau (R+|\nabla f|^{2})+f-n)dm, \\
\sigma  &=&2\tau ^{4}\int_{M}|R_{ij}+\nabla _{i}\nabla _{j}f-\frac{1}{2\tau }%
g_{ij}|^{2}dm{,}
\end{eqnarray*}%
where
\[
dm=udV,\qquad u=(4\pi \tau )^{-\frac{n}{2}}\mathrm{e}^{-f}.
\]

From the above formulas, we see that the MTS--fluctuation $\sigma
$ is nonnegative; it vanishes only on a gradient shrinking
soliton. $\left\langle E\right\rangle $ is
nonnegative as well, whenever the flow exists for all sufficiently small $%
\tau >0$. Furthermore, if the MTS--heat function $u$: (a) tends to
a $\delta -$function as $\tau \rightarrow 0,$ or (b) is a limit of
a sequence of partial heat functions $u_{i},$ such that each
$u_{i}$ tends to a $\delta -$function as $\tau \rightarrow \tau
_{i}>0,$ and $\tau _{i}\rightarrow 0,$ then the MTS--entropy $S$
is also nonnegative. In case (a), all the quantities $\left\langle
E\right\rangle ,S,\sigma $ tend to zero as $\tau \rightarrow 0,$
while in case (b), which may be interesting if $g_{ij}(\tau )$
becomes singular at $\tau =0,$ the MTS--entropy $S$ may tend to a
positive limit.

\section{Monoclonal antibodies for MTS--decay control}

To keep the MTS within the elliptic geometry with the positive
scalar curvature, $R>0$, which would enable the exponential decay
of its volume, we need the help from the local immune system.

Recall that monoclonal antibodies (mAb) are monospecific
antibodies that are identical because they are produced by one
type of immune cell that are all clones of a single parent cell.
Given (almost) any substance, it is possible to create monoclonal
antibodies that specifically bind to that substance; they can then
serve to detect or purify that substance \cite{Wiki}.

The idea of a `magic bullet' was first proposed by Paul Ehrlich
(Nobel Prize in Physiology or Medicine in 1908), who a century a
go postulated that if a compound could be made that selectively
targeted a disease-causing organism, then a toxin for that
organism could be delivered along with the agent of selectivity.

The invention of monoclonal antibodies is generally accredited to
Georges Köhler, César Milstein, and Niels Kaj Jerne in 1975
\cite{Kohler}, who shared the Nobel Prize in Physiology or
Medicine in 1984 for the discovery. The key idea was to use a line
of myeloma cells that had lost their ability to secrete
antibodies, come up with a technique to fuse these cells with
healthy antibody producing B--cells, and be able to select for the
successfully fused cells.

Human monoclonal antibodies are produced using transgenic mice or
phage display libraries. Human monoclonal antibodies are produced
by transferring human immunoglobulin genes into the murine genome,
after which the transgenic mouse is vaccinated against the desired
antigen, leading to the production of monoclonal antibodies. Phage
display libraries allow the transformation of murine antibodies in
vitro into fully human antibodies.

Antibody--directed enzyme prodrug therapy (ADEPT) involves the
application of cancer associated monoclonal antibodies which are
linked to a drug--activating enzyme. Subsequent systemic
administration of a non--toxic agent results in its conversion to
a toxic drug, and resulting in a cytotoxic effect which can be
targeted at malignant cells. The clinical success of ADEPT
treatments has been limited to date \cite{Francis}. However it
holds great promise, and recent reports suggest that it will have
a role in future oncological treatment \cite{Carter}.


\section{Appendix: Riemann and Ricci curvatures on a smooth $n-$manifold}

Recall that proper differentiation of vector and tensor fields on a smooth
Riemannian $n-$manifold is performed using the \textit{Levi--Civita
covariant derivative} (see, e.g., \cite{GaneshSprBig,GaneshADG}). Formally,
let $M$ be a Riemannian $n-$manifold with the tangent bundle $TM$ and a
local coordinate system $\{x^{i}\}_{i=1}^{n}$ defined in an open set $%
U\subset M$. The covariant derivative operator, $\nabla _{X}:C^{\infty
}(TM)\rightarrow C^{\infty }(TM)$, is the unique linear map such that for
any vector fields $X,Y,Z,$ constant $c$, and function $f$ the following
properties are valid:
\begin{eqnarray*}
\nabla _{X+cY} &=&\nabla _{X}+c\nabla _{Y}, \\
\nabla _{X}(Y+fZ) &=&\nabla _{X}Y+(Xf)Z+f\nabla _{X}Z, \\
\nabla _{X}Y-\nabla _{Y}X &=&[X,Y],
\end{eqnarray*}%
where $[X,Y]$ is the Lie bracket of $X$ and $Y$ (see, e.g., \cite{VladSiam}%
). In local coordinates, the metric $g$ is defined for any orthonormal basis
$(\partial_i=\partial_{x^i})$ in $U\subset M$ by
\begin{equation*}
g_{ij}=g(\partial_{i},\partial_{j})=\delta _{ij}, \qquad
\partial_{k}g_{ij}=0.
\end{equation*}
Then the affine \textit{Levi--Civita connection} is defined on $M$ by
\begin{equation*}
\nabla _{\partial _{i}}\partial _{j}=\Gamma _{ij}^{k}\partial _{k},\qquad
\text{where}
\end{equation*}
\begin{equation*}
\Gamma _{ij}^{k}=\frac{1}{2}g^{kl}\left( \partial _{i}g_{jl}+\partial
_{j}g_{il}-\partial _{l}g_{ij}\right)
\end{equation*}%
are the (second-order) \textit{Christoffel symbols}.

Now, using the covariant derivative operator $\nabla _{X}$ we can define the
\textit{Riemann curvature} $(3,1)-$tensor $\mathfrak{Rm}$ by (see, e.g.,
\cite{GaneshSprBig,GaneshADG})
\begin{equation*}
\mathfrak{Rm}(X,Y)Z=\nabla _{X}\nabla _{Y}Z-\nabla _{Y}\nabla _{X}Z-\nabla
_{\lbrack X,Y]}Z.
\end{equation*}%
$\mathfrak{Rm}$ measures the curvature of the manifold by expressing how
noncommutative covariant differentiation is. The $(3,1)-$components $%
R_{ijk}^{l}$ of $\mathfrak{Rm}$ are defined in $U\subset M$ by
\begin{eqnarray*}
\mathfrak{Rm}\left( \partial _{i},\partial _{j}\right) \partial _{k}
&=&R_{ijk}^{l}\partial _{l},\qquad\text{which expands as \cite{MTW},} \\
R_{ijk}^{l} &=&\partial _{i}\Gamma _{jk}^{l}-\partial _{j}\Gamma
_{ik}^{l}+\Gamma _{jk}^{m}\Gamma _{im}^{l}-\Gamma _{ik}^{m}\Gamma
_{jm}^{l}.
\end{eqnarray*}%
Also, the Riemann $(4,0)-$tensor $R_{ijkl}=g_{lm}R_{ijk}^{m}$ is defined as
the $g-$based inner product on $M$,
\begin{equation*}
R_{ijkl}=\left\langle \mathfrak{Rm}\left( \partial _{i},\partial _{j}\right)
\partial _{k},\partial _{l}\right\rangle .
\end{equation*}

The first and second Bianchi identities for the Riemann $(4,0)-$tensor $%
R_{ijkl}$ hold,
\begin{eqnarray*}
R_{ijkl}+R_{jkil}+R_{kijl} &=&0, \\
\nabla _{i}R_{jklm}+\nabla _{j}R_{kilm}+\nabla _{k}R_{ijlm} &=&0,
\end{eqnarray*}%
while the twice contracted second Bianchi identity reads
\begin{equation}
2\nabla _{j}R_{ij}=\nabla _{i}R.  \label{cB}
\end{equation}

The $(0,2)$ \textit{Ricci tensor} $\mathfrak{Rc}$ is the trace of the
Riemann $(3,1)-$tensor $\mathfrak{Rm}$,
\begin{equation*}
\mathfrak{Rc}(Y,Z)+\mathrm{tr}(X\rightarrow \mathfrak{Rm}(X,Y)Z),
\end{equation*}
so that
\begin{equation*}
\mathfrak{Rc}(X,Y)=g(\mathfrak{Rm}(\partial _{i},X)\partial _{i},Y),
\end{equation*}
Its components $R_{jk}=\mathfrak{Rc}\left( \partial _{j},\partial
_{k}\right) $ are given in $U\subset M$ by the contraction (see
e.g., \cite{MTW})
\begin{eqnarray*}
R_{jk} &=&R_{ijk}^{i},\qquad \text{or, in terms of Christoffel symbols,} \\
R_{jk} &=&\partial _{i}\Gamma _{jk}^{i}-\partial _{k}\Gamma
_{ji}^{i}+\Gamma _{mi}^{i}\Gamma _{jk}^{m}-\Gamma _{mk}^{i}\Gamma
_{ji}^{m}.
\end{eqnarray*}
Being a symmetric second--order tensor, $\mathfrak{Rc}$ has
$\QOVERD( ) {n+1}{2}$ independent components on an $n-$manifold
$M$. In particular, on a 3--manifold, it has 6 components, and on
a 2D surface it has only the following 3 components:
\[
R_{11}=g^{22}R_{2112},\qquad R_{12}=g^{12}R_{2121},\qquad
R_{22}=g^{11}R_{1221},
\]
which are all proportional to the corresponding coordinates of the
metric tensor,
\begin{equation}
\frac{R_{11}}{g_{11}}=\frac{R_{12}}{g_{12}}=\frac{R_{22}}{g_{22}}=-\frac{R_{1212}}{\det
(g)}. \label{Rsrf}
\end{equation}

Finally, the scalar curvature $R$ is the trace of the Ricci tensor $%
\mathfrak{Rc}$, given in $U\subset M$ by
\begin{equation*}
R=g^{ij}R_{ij}.
\end{equation*}

\end{document}